# Peculiar emission line generation from ultra-rapid quasi-periodic oscillations of exotic astronomical objects

E.F. Borra

Centre d'Optique, Photonique et Lasers

Département de Physique, Université Laval

Address: Département de Physique, Université Laval, Québec, Canada G1V 0A6

Email : borra@phy.ulaval.ca

**ABSTRACT**

The purpose of this article is to alert Astronomers, particularly those using spectroscopic surveys, to the fact that exotic astronomical objects (e.g. quasars or active galactic nuclei) that send ultra-rapid quasi periodic pulses of optical light would generate spectroscopic features that look like emission lines. This gives a simple technique to find quasi periodic pulses separated by times smaller than a nanosecond. One should look for emission lines that cannot be identified with known spectral lines in spectra. Such signals, generated by slower pulses, could also be found in the far infra-red, millimeter and radio regions, where they could be detected as objects unusually bright in a single narrow-band filter or channel. The outstanding interest of the technique comes from its simplicity so that it can be used to find ultra-rapid quasi-periodic oscillators in large astronomical surveys. A very small fraction of objects presently identified as Lyman $\alpha$ emitters that do not have other spectral features to confirm the Lyman $\alpha$ redshift, may possibly be quasi-periodic oscillators. However this is only a hypothesis that needs more observations for confirmation.

# 1. Introduction

The time domain is the least investigated of all the Astronomical domains (Fabian 2010). The short timescale domain is particularly difficult to investigate because one needs specialized instrumentation to obtain a very large quantity of data. On the other hand, astronomical objects that vary periodically with times shorter than nanoseconds could easily be detected by searching for periodic modulations in their spectra (Borra 2010).

Borra (2010) suggested a very simple technique to find objects that vary with periods shorter than nanoseconds. (Borra 2010) uses a Fourier transform analysis to demonstrate that pulses emitted with periodic constant time separations introduce spectral modulations that, although not periodic in wavelength units, are periodic in frequency units. Laboratory experiments (Chin et al. 1992) validate the theoretical analysis in Borra (2010). The outstanding interest of this spectroscopic technique comes from its simplicity since weak periodic modulations can easily be found by carrying out Fourier transform analyses of spectra converted to frequency units. Following the theoretical work in Borra (2010), a Fourier transform analysis of 2.5 million spectra from the Sloan Digital Sky Survey database was carried out to detect periodic modulations (Borra 2013). Periodic modulations were found in the spectra of only 223 galaxies of the 0.9 million galaxies observed. No periodic modulations were found in over 0.5 million spectra of stars and quasars. Most of the stars were significantly brighter than the galaxies so that there is no signal to noise ratio bias to favor galaxies. There is no doubt that the discovered periodic modulations are real, because the periods as a function of the redshifts of the host galaxies increase linearly with redshift along two very tight separate parallel lines. These linear increases with redshift indicate that the periods are the same in the rest frame of the galaxies with two base periods of $1.03\times10^{-13}$ and $1.09\times10^{-13}$ seconds. Borra (2013) concludes that they are probably caused by pairs of pulses separated by times of $1.03\times10^{-13}$ and $1.09\times10^{-13}$ seconds. Considering that the optical fibers used by the spectrograph of the Sloan Digital Sky Survey sample the center of the galaxies, we can assume that the light pulses are generated in the cores of the galaxies.

Some astronomical objects are known to have quasi-periodic oscillations that come from black hole binaries (see Done et al. 2007 and Remillard & McClintock 2006 for reviews). These objects are called Quasi-Periodic Oscillators (QPOs). High-frequency QPOs emit quasi-periodic oscillations with average oscillation frequencies as high as 500 Hz and therefore average periods of 2 milliseconds. Many theoretical models have been made but there is no consensus on the actual physical cause of the quasi-periodic oscillations. The introduction in Varniere & Vincent (2016) gives a brief summary of the many existing theoretical models.

Because some stellar QPOs, having periods of the order of 2 milliseconds have already been found, we can assume that some highly exotic objects, such as quasars and active galactic nuclei, may be QPOs capable of extremely rapid quasi-periodic oscillations having times shorter than nanoseconds. The hypothesis of the existence of ultra-rapid QPOs, that we shall abbreviates as URQPOs from here on, seems valid because Borra (2013) has found spectroscopic evidence that the nuclei of some galaxies generate pulses separated by times shorter than nanoseconds. We shall consider a simple method to find the signature of extremely rapid quasi-periodic pulsations in spectra. Note however that we are not trying to model known QPOs, therefore we do not assume exactly the same type of quasi-periodic oscillations.

2. Generation of spectral lines from ultra-rapid quasi-periodic oscillations

We start from the theoretical analysis in Borra (2010) that models the time variations of the electric field $E(t)$ emitted by an exotic astronomical source that sends $N$ periodic pulses of electric fields $V(t)$, separated by equal time intervals $\tau$, by convolving $V(t)$ with the comb function $\sum_{m} \delta(t - t_m)$, made by the sum of delta functions $\delta(t - t_m)$ separated by $t_m = (m-1)\tau$. with $m$ an integer number. $E(t)$ is then given by the convolution of $V(t)$ and $\delta(t - t_m)$

$$E(t) = V(t) \otimes \sum_{m} \delta(t - t_m) \quad . \tag{1}$$

One can obtain the spectral frequency dependence of such a signal by carrying out an analysis based on a Fourier transform of Equation 1. Starting from Equation 1, Borra (2010) finds that an astronomical quadratic detector (e.g. a CCD ) in a spectrograph detects a spectroscopic modulation signal given by

$$S(\omega) = S_1(\omega)\left[\sin(\omega N\tau/2)/(\sin(\omega\tau/2))\right]^2 , \qquad (2)$$

with $\omega = 2\pi\nu$ and $\nu$ the spectral frequency. As discussed in Borra (2010), a CCD measures the time average of $S(\omega)$, with the average carried over the entire time of observation. Note that we use the spectral frequency $\nu$ instead of the wavelength $\lambda$, that is usually used in Astronomy, because the spectral modulation is periodic in frequency units and therefore not in wavelength units. The maxima of $S(\omega)$ are separated by $\Delta\omega = 2\pi n/\tau$, with $n$ an integer number. Figure 1 in Borra (2010) shows the type of spectral signatures predicted by Equation (2). The shapes of the spectral signatures $S(\omega)$, measured with a spectrograph and a quadratic detector (e.g. a CCD), sharpen rapidly as the number of periodic pulses $N$ increases. This gives $S(\omega)$ a comb like shape with very sharp teeth for a very large number of periodic pulses. This paragraph only gives a small summary of the basic theory, see Borra (2010) for more details.

Following the analysis in Borra (2010) we can, for a very large number of periodic pulses separated by the constant period $\tau$, model the electric field signal as a function of time $E(t)$ by the convolution of a comb Shah function

$$III(t - n\tau) = \sum_n \delta(t - n\tau) , \qquad (3)$$

where $t$ is time, $\tau$ is the period of variation and $n$ is an integer number, with a Gaussian that models the shapes of the individual pulses. The Fourier transform of the Shah function $III(t-n\tau)$ gives a Shah function $III(\nu-n/\tau)/\tau$ with period $P = 1/\tau$ in the frequency domain $\nu$. The Fourier transform of the convolution of 2 functions gives the product of the Fourier transforms of the 2 functions. The Fourier transform of a Gaussian with a dispersion $\sigma$ gives another Gaussian with a dispersion $1/\sigma$. Therefore the Fourier transform of the electric field signal $E(t)$ made of pulses periodically spaced by $\tau$, modeled by the Shah function $III(t-n\tau)$ convolved with a Gaussian function with a dispersion $\sigma$, is the Shah function $III(\nu-n/\tau)/\tau$ multiplied by a Gaussian with a dispersion $1/\sigma$ centered at $\nu = 0$. The frequency spectrum is thus made of very narrow spectral lines, separated by a constant separation $\Delta \nu = 1/\tau$, having strengths that decrease with increasing frequency. In the case of quasi-periodic oscillations, the time separation $\tau$ is not constant and the frequency spectrum will not be a Shah function convolved with a Gaussian. Starting from this analytical discussion, we shall carry out numerical simulations of the Shah function where the separation between the pulses is no longer rigorously periodic but varies with random deviations from the period $\tau$.

We first model the time dependent electric field signal $E(t)$ of a periodic pulsator by the convolution of a Shah function with a Gaussian function that models the shapes of the pulses as a function of time. The spectral energy distribution, also called power spectrum, can be obtained from the Fourier transform of the autocorrelation of $E(t)$. This is discussed in many Optics textbooks (e.g. chapter 11 in Hecht 2002). Figure 1 shows the spectrum in frequency units obtained from the Fourier transform of the autocorrelation of the electric field signal $E(t)$ of a periodic pulsator that uses the model discussed in the previous paragraphs. The spectrum is in frequency units because, as explained in many textbooks, the Fourier transform of a signal in time units gives a spectrum in frequency units. Consequently, the Fourier transform of the autocorrelation of $E(t)$ gives a power spectrum in frequency units. The period is $1.67 \times 10^{-15}$ seconds and the width of the pulses at half-maximum intensity is $4.5 \times 10^{-16}$ seconds. Figure 1 validates the theoretical discussion based on the Shah function in the previous paragraph. The spectrum in Figure 1 shows the periodically separated emission lines of decreasing strength that are expected from a Shah function multiplied by a Gaussian. Computer simulations of the Chin et al. (1992) experiment were also carried out to further validate our numerical analysis. Figure 2 shows the simulated spectrum of the recombed pulses in a Michelson interferometer for the case where there are only 2 pulses. It shows the type of signal in the Chin et al. (1992) experiment, as can be seen by comparing our Figure 2 to their figure 1, thereby validating our numerical simulations. Note that Figure 1 gives the entire spectrum as it would be seen by a spectrograph capable of observing the entire spectrum, while Figure 2 gives the spectrum generated by the recombined pulses in a Michelson interferometer in the Chin et al. (1992) experiment. This explains why the spectrum is not symmetrical and centered at a frequency of 0.0 in Figure 1, while the spectrum in Figure 2 is symmetrical and is centered at a particular location given by the path difference in the interferometer.

To understand the effect of degrading the periodicity by changing the periodic time locations of the pulses, numerical simulations were carried out with matlab software. They start from the same model of *E(t)* obtained from the convolution of a Shah function with a Gaussian that generated Figure 1, but use software that changes at random the time locations of the individual periodic pulses predicted by the constant time separation $\tau$ in the Shah function. They use the same value of the period $\tau$ and the same dispersion of the Gaussian that models the individual pulses used to generate Figure 1. The time locations of the pulses are changed with the matlab random number generator function *rand* that generated numbers *r* that vary at random between *r = 0.0* and *r = 0.5*. These random numbers *r* then multiply the equal time separation $\tau$ in the Shah function to generate a separation $\Delta\tau = r*\tau$ from the basic period $\tau$, that varies at random among the pulses. The $\Delta\tau$ separation from the periodicity $\tau$ among the pulses therefore varies at random between $\Delta\tau = 0.0*\tau$ and $\Delta\tau = 0.5*\tau$. A different random time deviation is therefore added to the time location of every one of the peaks of the pulses predicted by the Shah function model. Figure 3 shows the pulsating quasi-periodic *E(t)* signal generated with this numerical technique. Like the signal used to generate Figure 1, the period is $1.67 \times 10^{-15}$ seconds and the width of the pulses at half-maximum intensity is $4.5 \times 10^{-16}$ seconds. The intensity pulses *I(t)* are given by the time average, set by the detector, of the square of *E(t)* and therefore have the same locations and similar shapes. Figure 4 shows the spectrum in frequency units generated by this type of signal. Comparing Figure 4 to Figure 1, we see that the relative strength of the emission lines in Figure 1, which was generated with the same period $\tau$ but with $\Delta\tau = 0.0$ so that the pulses are rigorously periodic, has changed significantly. The first emission line in Figure 1 totally dominates in Figure 4 and the other emission lines have become extremely weak. There is also a weak noisy spectral background that comes from the energy lost from the emission lines in Figure 1. In an actual astronomical spectrum there would also be a spectral background (e.g. from a host galaxy), as well as noise, and the spectral signature of such a URQPO would be interpreted as being an emission line.

A peculiar feature in comparing Figure 4 to Figure 1 comes from the fact that, in Figure 4, although very weak, the emission line at $1.810^{15}$ Hz can be seen while the emission line at $1.210^{15}$ Hz cannot be seen. Intuitively one would expect the emission line at $1.210^{15}$ Hz to be stronger than the emission line at $1.810^{15}$ Hz in Figure 4 because the emission line at $1.210^{15}$ Hz is 3.3 times stronger than the emission line at $1.810^{15}$ Hz in Figure 1. This peculiarity comes from the computer simulations that generated Figure 4 and is simply due to the fact that, as discussed in section 2, they change at random the time locations of the individual periodic pulses. This adds noise in Figure 4 and changes the relative strength of the emission features in a noisy manner, thereby explaining why, in Figure 4, the emission line at $1.810^{15}$ Hz can be seen, while the emission line at $1.210^{15}$ Hz cannot be seen.

The intensities in Figures 1 to 4 are in arbitrary units and one may wonder about the energy required to generate an emission line that is detectable in an astronomical spectrum. This is a highly complex issue, since the required energy depends on many parameters, like the relative strength with respect to the continuum, the width of the line and whether the emission is isotropic or within a jet. The emission is likely to occur in a jet, in which case the total energy emitted will also depend on the width of the jet and its angular distribution. On the other hand, the energy and, particularly, the energy density required are important issues that are briefly discussed in section 3 as well as in Borra (2010).

3. Discussion

The main conclusion of the simulations can be seen in Figure 4 that shows a very strong emission line followed by much weaker emission lines that would be buried in noise in astronomical spectra. This URQPO signal would therefore be interpreted as a single emission line. The emission line in Figure 4 has a very narrow width and would therefore have, in a spectrum, the width of the instrumental profile of the spectrograph.

On the other hand this emission line is generated by a signal having a constant average period $\tau$ to which one adds a deviation $\Delta\tau$ that varies at random among the pulses. The value of the period $\tau$ determines the location of the emission line in the spectrum since its location is given by $v = 1/\tau$. If we consider an URQPO that has an averaged period $\tau$ that varies slowly in time, to which one adds random variations $\Delta\tau$, one would obtain a broader emission line, since one would add many lines similar to the single line in Figure 4 but at different frequency locations and with different relative intensities. This is the type of signal that could be generated, for example, by sending many flashes of light lasting a few seconds, each one made of a large number of quasi-periodic pulses having average periods $\tau$ that vary at random among the flashes. The spectrograph would therefore detect a wider emission line having a shape that depends on the variations of $\tau$ within the total time of observation used. Note also that for an URQPO that does not have random variations as strong as those assumed to generate Figure 4. the other emission lines in Figure 1 would not be as much weakened and may therefore be detectable.

One may wonder what kind of physical phenomenon might produce quasi-periodic pulses. This is a difficult question to answer because, to begin with, the physics responsible for quasi-periodic oscillations in presently discovered QPOs is not understood (Varniere & Vincent 2016). An easy answer is that, in principle, any peculiar physical phenomenon can occur in exotic objects like black holes. This hypothesis is validated by the fact that, as mentioned in the introduction, binary black holes generate quasi-periodic oscillations. Although the known QPOs have much longer time scales than those that we assume, the discovery of spectral modulations probably generated by pulses separated by times of the order of $10^{-13}$ seconds (Borra 2013) validates the hypothesis of quasi-periodic oscillations having very small time scales. One may of course wonder whether the time scales of $\tau$ will increase or decrease with the mass of the black hole. It is presently impossible to answer this question because, firstly, the physical cause of the oscillations in known QPOs is totally unknown and, secondly, the hypothetical physical cause in very massive URQPOs could be totally different of the physical cause in the known QPOs that have much smaller masses. Black holes are poorly understood exotic objects and we can assume anything. Furthermore, the hypothetical quasi-periodic pulses may not come from black hole oscillations, but be emitted in jets. The analysis in Borra (2013) concludes that the detected signals are generated in jets.

We have only considered the signatures of URQPOs in spectra; however slower URQPOs, that have larger values of the basic period $\tau$, would emit a signal that may be detected in filters or channels at lower frequencies. One should therefore look for such a signal in the far infra-red, millimeter and radio regions.

At first sight, the energy, and the energy density requirements, that come from the required short timescales seem incompatible with physics. However exotic objects may use such exotic physics. This hypothesis is validated by the fact that some known astronomical sources also have major energy density problems that are difficult to understand. For example, Hankins et al. (2003) have detected nanosecond radio pulses and Hankins & Eilek (2007) discuss observed nanopulses that are unresolved at a time resolution of 0.4 ns and imply brightness temperatures of $2 \times 10^{41}$ K, which imply energy densities $10^{136}$ times the energy density at the center of the Sun. The energy considerations of ultra-rapid pulses are discussed in greater details in Borra (2010).

To generate our figures, we use a very simple model consisting of the convolution of a Shah function to model a periodic electric field as a function of time signal with a Gaussian function to model the shapes of the individual pulses. Obviously, a hypothetical URQPO is likely to generate more complex signals. However, this would not affect our main conclusions. More complex shapes of the pulses would simply multiply the Shah function *III( ν-n/τ)/τ* with the Fourier transform of the actual shape of the pulses instead of the Fourier transform of the Gaussian model. Assuming pulses that vary in intensity and shape also does not affect our conclusions. While the peaks of the *I(t)* signal used to generate Figure 4 have equal amplitudes (see Figure 3), simulations carried out with peaks that varied at random in amplitude gave results similar to those in Figure 4. Varying pulse shapes would simply affect the average shape of the function that multiplies the Fourier transform Shah function. The purpose of the simulation is simply to show the effect of very rapid quasi-periodic oscillations in spectra.

## 4. Conclusion

Ultra-rapid pulsators that emit quasi-periodic pulses having time separations smaller than $10^{-10}$ seconds generate a signal that would be identified as an emission line in astronomical spectra. One should therefore look for emission lines that cannot be identified as known spectral lines.

URQPOs are probably very rare objects and the best way to find URQPOs is in large astronomical surveys (e.g. the Sloan Digital Sky Survey) where one should look for objects with a single emission line and then take high quality spectra to verify whether other spectral features confirm, or deny, the hypothesis that the emission line originate from an URQPO.

Strong emission lines have been detected in some faint astronomical objects (Hu, Cowie & McMahon 1998) that are called Lyman $\alpha$ emitters. They have been extensively studied over the past 20 years and the majority of them are clearly Lyman $\alpha$ emitters since they have other spectral features (e.g. a CIV line or a Lyman break) that confirm the redshift of the Lyman $\alpha$ line. It is however possible that a very small fraction of Lyman $\alpha$ emitters are actually Quasi-Periodic Oscillators. This could be the case for objects classified as Lyman $\alpha$ emitters that do not have other spectral features that confirm the Lyman $\alpha$ redshift. Note however that we are not suggesting that Lyman $\alpha$ emitters are URQPOs, but only that it is possible that a very small fraction of them are URQPOs. We only make this hypothesis because Lyman $\alpha$ emitters have the type of strong emission line expected from an URQPO. Note that two of the Lyman $\alpha$ emitters in figure 1 in Hu, Cowie & McMahon (1998) only show a single strong emission line. They therefore have the kind of URQPO signal predicted in this work.

To verify whether the emission line in a spectrum is due to quasi-periodic oscillations or is a Lyman $\alpha$ line, one should look for spectral features that should be present in a Lyman alpha emitter. Because the candidates will be faint objects one should obviously look for strong spectral features. The strongest expected features are the Lyman series (e.g. Lyman $\alpha$ at 103.6 nm and the Lyman-break at 91.18 nm) at the redshift given assuming that the emission line is Lyman $\alpha$ Some other spectral features may also be present, but their strengths vary among Lyman α emitters (Cowie, Barger & Hu, 2010.

One could also find an URQPO signal in the far infrared, millimeter and radio regions, where an URQPO could be detected as being significantly brighter in a single narrow-band filter or channel among many filters or channels.

The spectral line in Figure 4 is very strong and one could detect such a strong line in the case where the signal from the pulses dominates the background spectrum. However the line would appear much weaker in the case where there is a strong background continuous spectrum coming from the host galaxy. It would also be broader than in Figure 4 if generated by many flashes, each flash containing many quasi-periodic pulses having an average period that varies among the flashes.

The possibility that extremely rapid URQPOs may actually exist is supported by the fact that some stellar QPOs have already been found. Although the known stellar QPOs are much slower than the hypothetical extremely rapid URQPOs, the existence of URQPOs is supported by the fact that Borra (2013) found periodic spectral modulations in the spectra of the cores of 223 galaxies. These spectral modulation, generated by ultra-rapid light variations, were predicted by Borra (2010) .

*Acknowledgements.* This research has been supported by the Natural Sciences and Engineering Research Council of Canada

References

Borra, E. F. 2010, A&A, 511, L6
Borra, E. F. 2013, ApJ, 774, 142
Chin, S. L., Francois, V., Watson, J. M., & Delisle, C. 1992, ApOpt, 31, 3383

Done , C., Gierlinsky, M,, & Kubota, A. 2007, A&A Rev., 15, 1

Fabian, A. C. 2010, in Serendipity: The Darwin College Lectures, Serendipity in Astronomy, ed. M. de Rond, I. M. de Rond, & I. Morley (Cambridge: Cambridge Univ. Press), 73

Hecht, E. 2002, Optics, (San Francisco, Addison Wesley)

Hu, E. M., Cowie, L. L., & McMahon, R. G. 1998, ApJ, 502. L99

Hankins, T. H., & Eilek, J. A. 2007, ApJ,, 670, 693

Hankins, T. N., Kern, J. S., Weatherall, J. C., & Ellek, J. A. 2003, Nature, 422, 141

Remillard, R.A., & McClintock, J,. E. 2006, ARA&A, 44, 49

Varniere, P., & Vincent, F.H. 2016, A&A, 591, A36

FIGURES

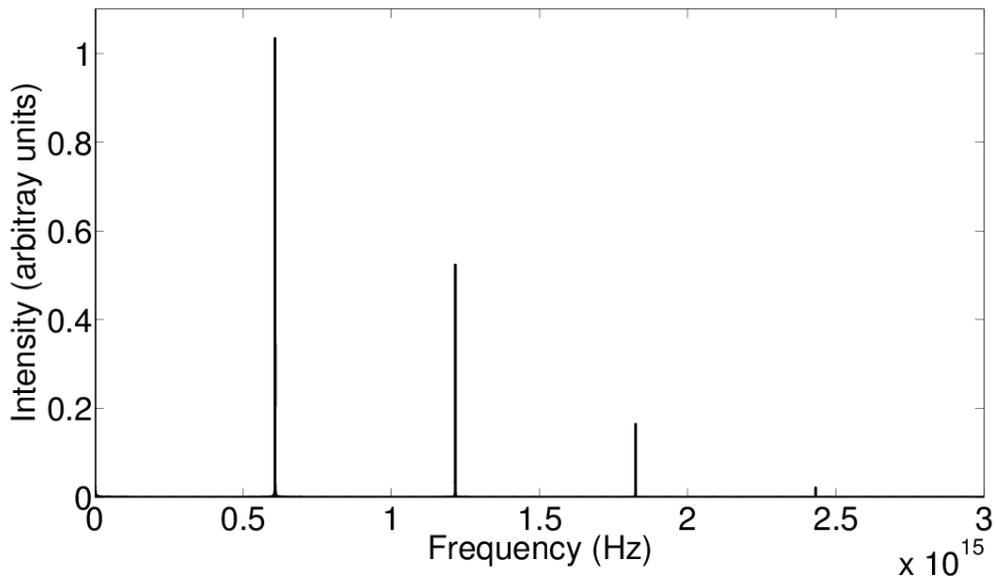

Figure 1

Spectrum in frequency units obtained from a periodic electric field signal *E(t)* modelled by a Shah function convolved with a Gaussian that models the shapes of the individual pulses. It validates the theoretical discussion based on the Shah function in section 2.

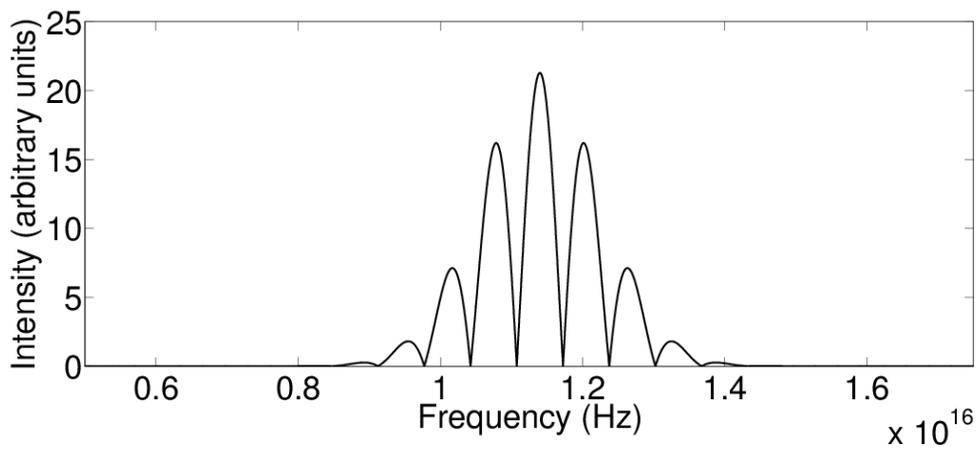

Figure 2

Spectrum in frequency units of the recombined pulses in a Michelson interferometer for the case where there are only 2 pulses in *E(t)*. It shows the same type of spectrum, obtained in the laboratory experiment carried out by Chin et al. (1992).

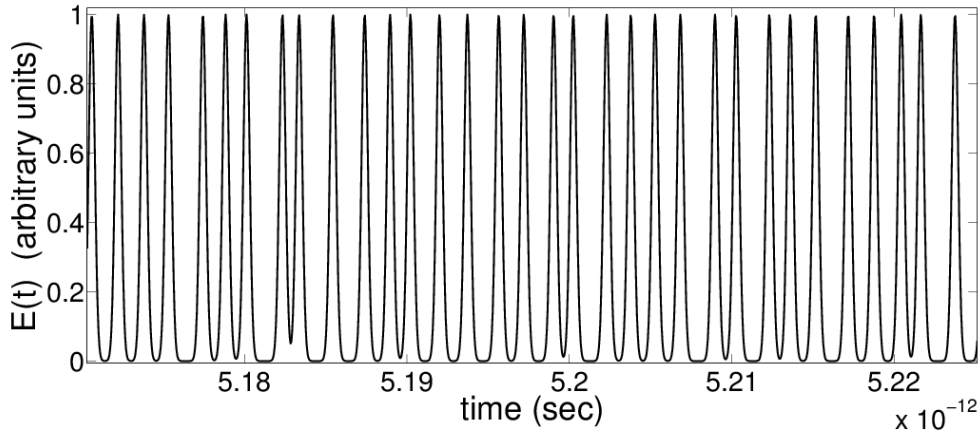

Figure 3

Part of the pulsating quasi-periodic electric field signal *E(t)* used to generate the URQPO spectrum in Figure 4. The *E(t)* signal starts from the same model used to generate Figure 1 but uses software that changes at random the time locations of the periodic pulses predicted by the constant time separation $\tau$ used by the *E(t)* signal that generated Figure 1. The $\Delta\tau$ separation from the average periodicity $\tau$ among the pulses varies at random between $\Delta\tau = 0.0*\tau$ and $\Delta\tau = 0.5*\tau$.

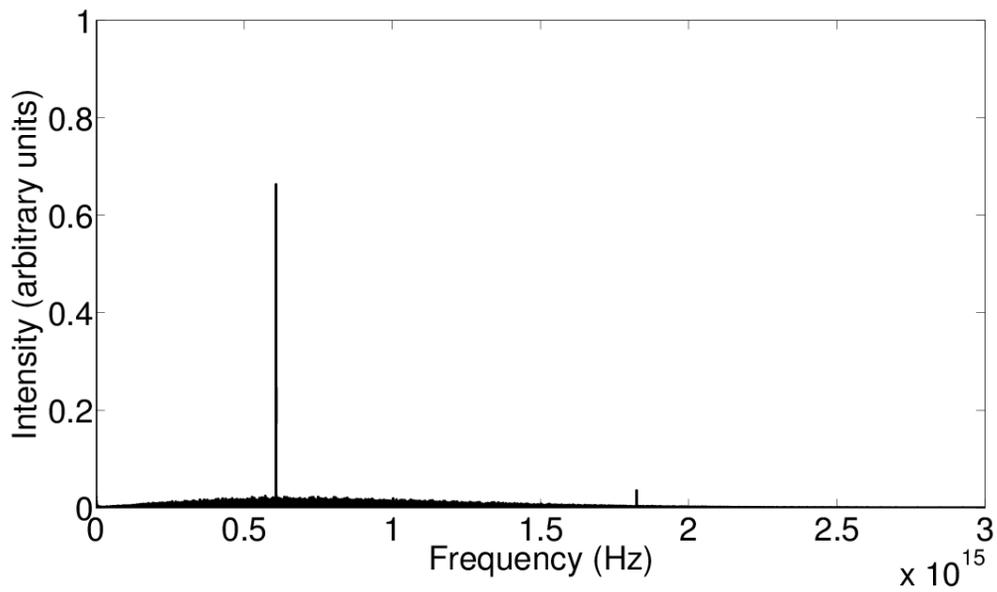

Figure 4

Spectrum in frequency units generated by the quasi-periodic electric field signal *E(t)* in Figure 3. It is dominated by a very strong narrow emission line followed by much weaker emission lines.